\documentclass[apj]{emulateapj}

\usepackage{color}

\shorttitle{Fluorene clusters photo-chemistry}
\shortauthors{Zhang et al.}

\begin{document}
\title{Laboratory photo-chemistry of covalently bonded fluorene clusters: observation of an interesting PAH bowl-forming mechanism}
\author{Weiwei Zhang$^{1,2,a}$, Yubing Si$^3$, Junfeng Zhen$^{1,2,*}$, Tao Chen$^{4,6}$, Harold Linnartz$^5$, Alexander G. G. M. Tielens$^{6}$} 

\affil{$^{1}$CAS Key Laboratory for Research in Galaxies and Cosmology, Department of Astronomy, University of Science and Technology of China, Hefei 230026, China} 
\affil{$^{2}$School of Astronomy and Space Science, University of Science and Technology of China, Hefei 230026, China}
\affil{$^{3}$Henan Provincial Key Laboratory of Nanocomposites and Applications, Institute of Nanostructured Functional Materials, Huanghe Science and Technology College, Zhengzhou 450006, China}
\affil{$^{4}$School of Engineering Sciences in Chemistry, Biotechnology and Health, Department of Theoretical Chemistry \& Biology, Royal Institute of Technology, 10691, Stockholm, Sweden}
\affil{$^{5}$Sackler Laboratory for Astrophysics, Leiden Observatory, Leiden University, P.O. Box 9513, 2300 RA Leiden, The Netherlands}
\affil{$^{6}$Leiden Observatory, Leiden University, P.O.\ Box 9513, 2300 RA Leiden, The Netherlands}   
\email{jfzhen@ustc.edu.cn}

\begin{abstract}
	
The fullerene C$_{60}$, one of the largest molecules identified in the interstellar medium (ISM), has been proposed to form top-down through the photo-chemical processing of large (more than 60 C-atoms) polycyclic aromatic hydrocarbon (PAH) molecules. In this article, we focus on the opposite process, investigating the possibility that fullerenes form from small PAHs, in which bowl-forming plays a central role. We combine laboratory experiments and quantum chemical calculations to study the formation of larger PAHs from charged fluorene clusters. The experiments show that with visible laser irradiation, the fluorene dimer cation - [C$_{13}$H$_{9}$$-$C$_{13}$H$_{9}$]$^+$ - and the fluorene trimer cation - [C$_{13}$H$_{9}$$-$C$_{13}$H$_{8}$$-$C$_{13}$H$_{9}$]$^+$ - undergo photo-dehydrogenation and photo-isomerization resulting in bowl structured aromatic cluster-ions, C$_{26}$H$_{12}$$^+$ and C$_{39}$H$_{20}$$^+$, respectively. To study the details of this chemical process, we employ quantum chemistry that allows us to determine the structures of the newly formed cluster-ions, to calculate the hydrogen loss dissociation energies, and to derive the underlying reaction pathways. These results demonstrate that smaller PAH clusters (with less than 60 C-atoms) can convert to larger bowled geometries that might act as building blocks for fullerenes, as the bowl-forming mechanism greatly facilitates the conversion from dehydrogenated PAHs to cages. Moreover, the bowl-forming induces a permanent dipole moment that - in principle - allows to search for such species using radio astronomy.

{
	\renewcommand{\thefootnote}%
	{\fnsymbol{footnote}}
	\footnotetext[1]{$^a$Current address: Department of Mechanical and Nuclear Engineering, Pennsylvania State University, University Park, PA 16802, United States.}
} 

\end{abstract}


\section{Introduction}
\label{sec:intro}

Progress in observational techniques, both ground based and from space missions, shows that molecular complexity in space may be beyond our imagination. Fullerenes, like C$_{60}$, are seen in very different environments and thought to be chemically linked to PAHs, as discussed in \citet{tie13}. Interstellar PAHs and PAH derivatives (e.g., PAH clusters) are believed to be very ubiquitous in the ISM, where they are generally thought to be responsible for the strong mid-infrared (IR) features in the 3-17 $\mu$m range that dominate the spectra of most galactic and extragalactic sources \citep{all89,pug89,sel84,gen98}. PAH cations have been proposed as carriers of the diffuse interstellar bands (DIBs) \citep{sal96,ger11}. IR spectra of circumstellar and interstellar sources have revealed the presence of the fullerenes C$_{60}$ and C$_{70}$ in space \citep{cam10, sel10}. Recently, several DIBs around 1 $\mu$m have been linked to electronic transitions of C$_{60}$$^+$ \citep{cam15, wal15, cor17}. Hence, understanding PAH and fullerene formation and destruction processes has attracted much attention in the field of molecular astrophysics \citep{tie13}.

Based on IR observations of interstellar reflection nebulae, Bern\'e et al. proposed that PAHs can be converted into graphene and subsequently to C$_{60}$ by photochemical processing combining the effects of dehydrogenation, fragmentation and isomerization \citep{berne12, berne2015}. This idea is supported by laboratory studies, which demonstrate that C$_{66}$H$_{22}$$^+$ can be transformed into C$_{60}$$^+$ upon irradiation, following full dehydrogenation, graphene flake folding and C$_2$ losing chemical pathways \citep{zhen2014b}. The conversion of graphene flakes to cages and fullerenes has been studied using transmission electron microscopy and quantum chemistry \citep{chu10}. \citet{pie14} have elucidated relevant reaction routes and molecular intermediaries in the conversion of graphene flakes into cages and fullerenes.

The photochemical breakdown of PAHs to smaller hydrocarbons and the conversion to fullerenes may be counteracted by a PAH-growth process that converts clusters of small PAHs into larger, fully aromatic PAHs. Such a process could start with the formation of van der Waals-bonded or charge transfer bonded PAHs clusters that are photochemically converted into large PAHs and PAH cations \citep{zhen2018}. The presence of PAH clusters in the deeper zones of photodissociation regions (PDRs) has been inferred from singular value decomposition analysis of ISO and Spitzer spectral maps of PDRs \citep{rapacioli05a, berne07}. \citet{rhee07}, on the other hand, suggest that PAH clusters are present near the surfaces of PDRs. Their suggestion is based upon a putative association of the Extended Red Emission \citep{vij04} with luminescence by charged PAH clusters. The formation and destruction of PAH clusters has been studied by \citep{rapacioli06}. In their analysis, these authors balanced coagulation with UV-driven photo-evaporation of weakly bonded van der Waals clusters. However, rather than thermal evaporation, photochemical evolution of van der Waals clusters may result in covalent bond formation. Experimental and ab-initio molecular dynamics studies on ionization of van der Waals bonded acetylene clusters reveal a reaction channel towards the benzene cation. In this route, the excess photon and chemical energy is taken away by an H-atom or by one of the spectator acetylenes in the cluster \citep{sti17}. Similar, UV-driven reaction routes may exist for PAH clusters resulting in covalently bonded PAH dimers, trimers or larger multimers. The presence of aromatic structures bonded by aliphatic links in space has also been postulated by \citet{mic12} as an explanation for spectral detail of the aromatic infrared bands. In addition, the driving force of the cluster growth in the soot nuclei formation process is the generation of PAHs radical, which is formed by releasing hydrogen atoms from PAHs. Afterwards, the reaction between PAH radicals and PAHs leads to covalently bonded clusters \citep{lep01, ric00,ric05}. Finally, we point towards experimental studies on the interaction of energetic ions with PAH clusters which also lead to covalently bonded PAH-multimers \citep{zet13, gat15, gat16}. 

In our experimental set-up, we have identified an efficient clustering method that naturally leads to covalently-bonded dimer and trimer cations. In an earlier study, we reported the photochemical conversion of such pyrene clusters into large, fully aromatic PAHs \citep{zhen2018}. Here, we extend these studies to fluorene clusters and demonstrate that the resulting molecules incorporate pentagons. Inclusion of pentagons in the molecular structure leads to curvature of the species. We have selected fluorene (C$_{13}$H$_{10}$) as a prototypical molecule for its unique molecular property: a carbon skeleton, in which two six-membered rings are fused to a central pentagonal ring. While its abundance in the ISM is unknown, fluorene has been detected in carbonaceous meteorites \citep{sep02}. In addition, fluorene is very amenable for experimental studies because of its small size. Likely, interstellar PAH clusters consist of PAHs that are much larger than fluorene. Nevertheless, the photochemical behavior of fluorene identifies a number of key processes that may play a role in the evolution of PAH clusters in space. Specifically, our results show that bowl-forming is an important aspect of the photochemical evolution of PAHs and this can be a first step in the formation of cages and fullerenes. 

As our experiments are not optimized to study covalent-bonded cluster formation, we focus here exclusively on the subsequent evolution under irradiation. Further studies – such as the acetylene studies \citep{sti17} –are required to address the kinetics of the formation of such clusters. Together with our experiment, such kinetic parameters will be necessary to evaluate the abundance of covalently bonded PAH clusters and their role in space. 

\section{Experimental Methods}
\label{sec:exp}

The experiments have been performed on i-PoP, our instrument for photo-dissociation of PAHs, a high vacuum ion trap time-of-flight (TOF) system that has been described in detail in \citep{zhen2014a}. In short, neutral fluorene (Aldrich, 98 \%) species are transferred into the gas phase from an oven ($\sim$ 305 K). The fluorene molecules are ionized by an electron gun. A steel mesh (hole diameter $\sim$ 0.1 mm) is put on top of the oven to increase the local density of fluorene molecules to facilitate cluster formation. The increased density in the fluorene plume allows cation cluster-formation as discussed in \citep{zhen2018}. In our experiments, the low energy barrier for H-loss from fluorene (2.3 eV, \citet{west18}) leads to efficient radical formation upon electron gun irradiation and promotes covalently bonded cluster formation. Once formed, cation species are transported into a quadrupole ion trap and trapped. The trapped ions are then irradiated by several (typically $\sim$ 5) pulses from a pulsed Nd:YAG pumped dye laser system, providing visible light (with wavelengths around 595 nm, linewidth $\sim$ 0.2 cm$^{-1}$, pulse duration $\sim$ 5 ns). After irradiation the trap is opened and dissociation products are measured using the TOF mass spectrometer.

\section{Experimental results and discussion}
\label{sec:results}

Typical TOF mass spectra of the fluorene dimer and trimer cations are shown in Figures 1 and 2, respectively. In the sample condition without laser irradiation (Figure 1, middle panel), several peaks, corresponding to different species, are observed for the fluorene dimer cation, i.e., these species are the direct result of the electron impact ionization. In order to interpret these mass peaks correctly, it is important to note that the chance of producing a $^{13}$C-containing cluster increases with the number of C-atoms involved. The method to discriminate for pure $^{12}$C and $^{13}$C polluted mass signals has been described in our previous studies \citep{zhen2014a}. After correction for the $^{13}$C isotope and normalization, we conclude that six dimer species are formed at this stage, namely C$_{26}$H$_{19}$$^+$ (0.2 \%), C$_{26}$H$_{18}$$^+$ (26.5 \%), C$_{26}$H$_{17}$$^+$ (41.6 \%), C$_{26}$H$_{16}$$^+$ (7.1 \%), C$_{26}$H$_{15}$$^+$ (19.6 \%), and C$_{26}$H$_{14}$$^+$ (5.0 \%). We emphasize that this procedure leads exclusively to dehydrogenated fluorene dimers. It is tempting to speculate that any (ionized) cluster that did not make a covalent bond and is bonded solely by weak van der Waals bonds will rapidly evaporate before or in the trap due to the excess internal energy.

Upon laser irradiation (Figure 1, lower panel, 0.3 mJ), many new peaks are observed in the mass spectrum (Figure 1, upper panel, differential spectrum). The dehydrogenation sequence differs from that observed for monomers such as hexa-peri-hexabenzocoronene (C$_{42}$H$_{18}$, HBC, \citet{zhen2014a}). The fragmentation pattern of HBC is characterized by even-H peaks stronger than odd-H peaks. In contrast, in the fragmentation pattern of the fluorene dimers, two short but different dehydrogenation sequences can be distinguished in the range of m/z=326$-$330 and m/z=318$-$326, respectively. In the range of m/z=326$-$330, the fraction of species with an odd number of hydrogens is higher than that with an even number, e.g., the intensity of C$_{26}$H$_{17}$$^+$ (m/z=329) and C$_{26}$H$_{15}$$^+$ (m/z=327) is stronger than that of their neighbor, C$_{26}$H$_{16}$$^+$ (m/z=328). On the other hand, in the m/z=318$-$326 range, the dehydrogenation behavior is that of regular PAHs with stronger even peaks than odd peaks, e.g., the intensity of the C$_{26}$H$_{12}$$^+$ (m/z=324) peak is stronger than that of its neighbors (C$_{26}$H$_{11}$$^+$, m/z=323 and C$_{26}$H$_{13}$$^+$, m/z=325). Following our pyrene dimer study, we suggest that the two different dehydrogenation patterns in these two m/z regions imply that the fluorene dimer cations (e.g., [C$_{13}$H$_{9}$$-$C$_{13}$H$_{9}$]$^+$ or C$_{26}$H$_{18}$$^+$) convert to aromatic species with more conjugated $\pi$-bonds (e.g., {C$_{26}$H$_{12}$$^+$}) after dehydrogenation. We will discuss details in the next section.

The typical mass spectrum of the fluorene trimer cluster cation is shown in Figure 2. Similar to the fluorene dimer mass region, without laser irradiation (Figure 2, middle panel), the dominant species are the partially dehydrogenated fluorene trimer cluster cations (e.g., C$_{39}$H$_{26}$$^+$ with m/z=494). With laser irradiation (Figure 2, lower panel, 0.4 mJ), a wide range of fragment ions is evident in the mass spectra (Figure 2, upper panel, differential spectrum). In this case no -H/-2H intensity alternations are found.

\begin{figure}[t]
	\centering
	\includegraphics[width=\columnwidth]{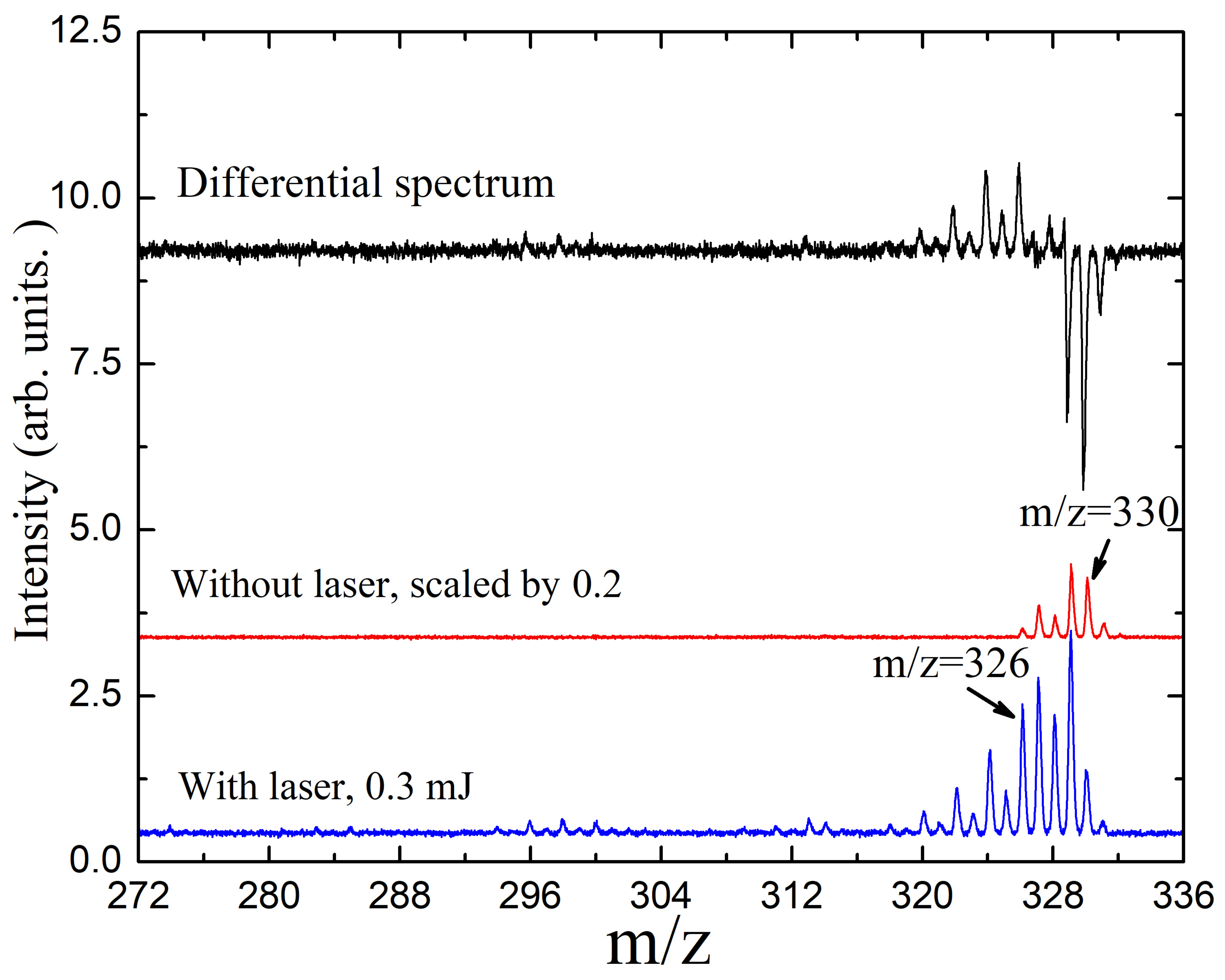}
	\caption{Mass spectrum of fluorene dimer cluster cations (e.g., C$_{26}$H$_{18}$$^+$, m/z=330) without irradiation (red), irradiated at 595 nm (blue) and the differential spectrum (black). }
	\label{fig1}
\end{figure}

\begin{figure}[t]
	\centering
	\includegraphics[width=\columnwidth]{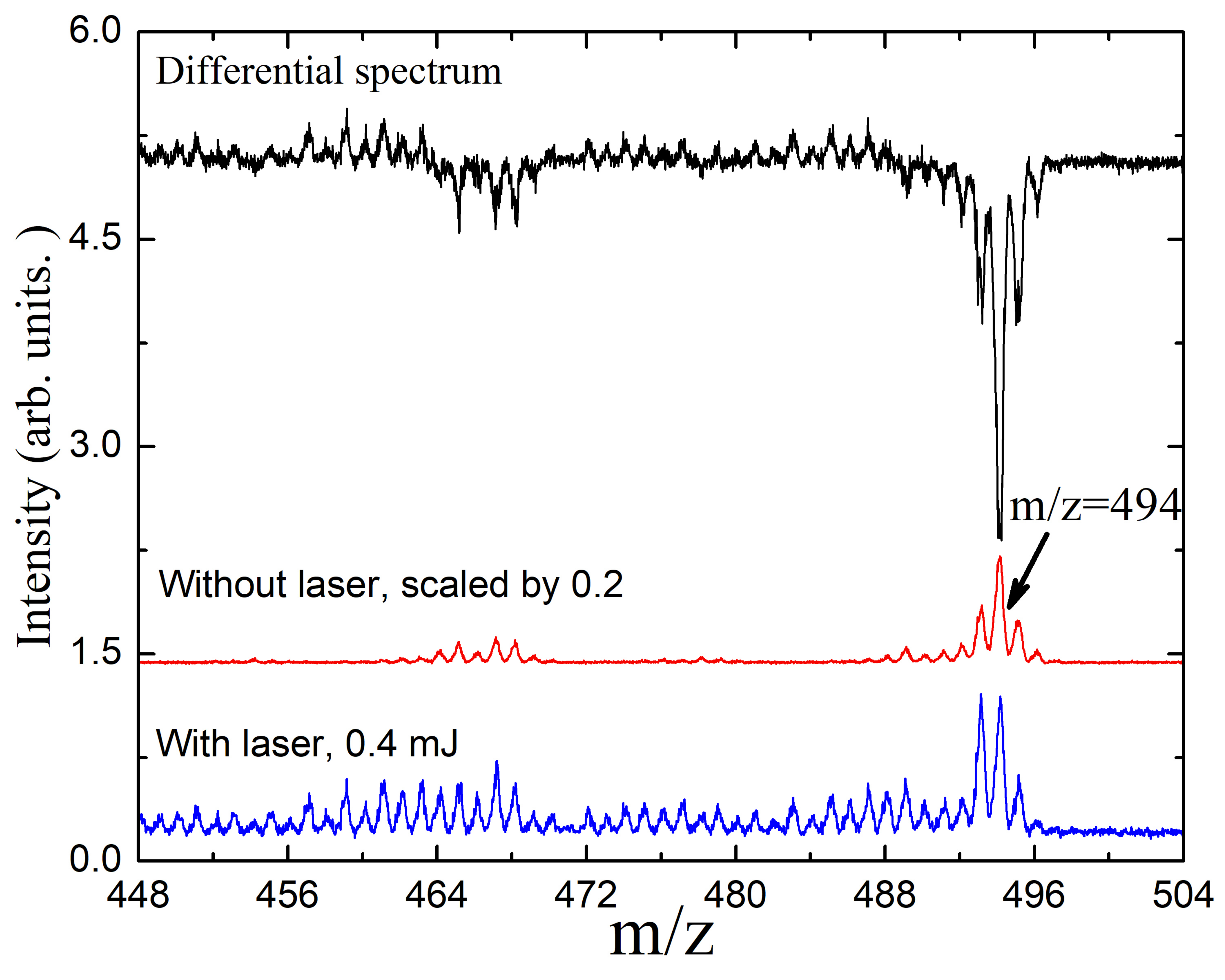}
	\caption{Mass spectrum of fluorene trimer cluster cations (e.g., C$_{39}$H$_{26}$$^+$, m/z=494) without irradiation (red), irradiated at 595 nm (blue) and the differential spectrum (black).}
	\label{fig2}
\end{figure}

\begin{figure}[t]
	\centering
	\includegraphics[width=\columnwidth]{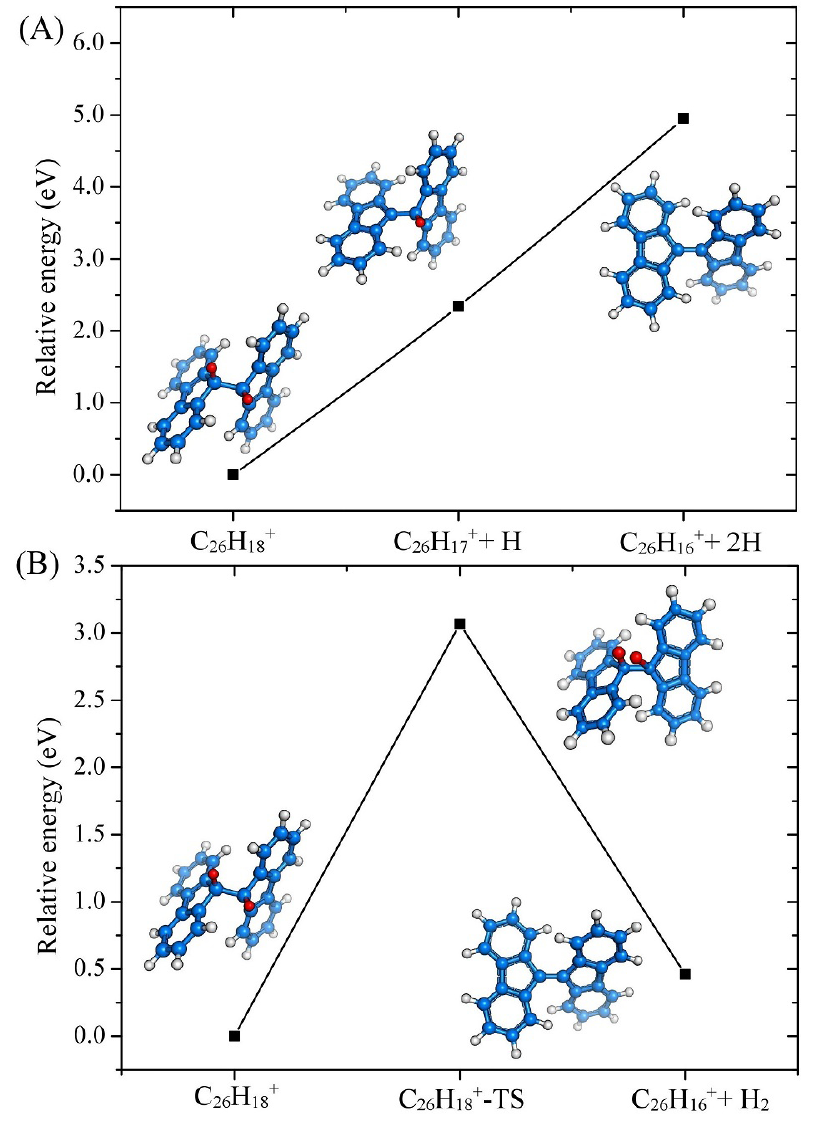}
	\caption{The reaction pathway of fluorene dimer cluster cation: panel (A) from C$_{26}$H$_{18}$$^+$ to C$_{26}$H$_{16}$$^+$ with H $+$ H loss channel; and panel (B) from C$_{26}$H$_{18}$$^+$ to C$_{26}$H$_{16}$$^+$ with H$_2$ loss channel. Carbon is shown in blue, nonreactive hydrogen in gray, and abstracted hydrogen in red.}
	\label{fig3}
\end{figure}

\begin{figure*}[t]
	\centering
	\includegraphics[width=\textwidth]{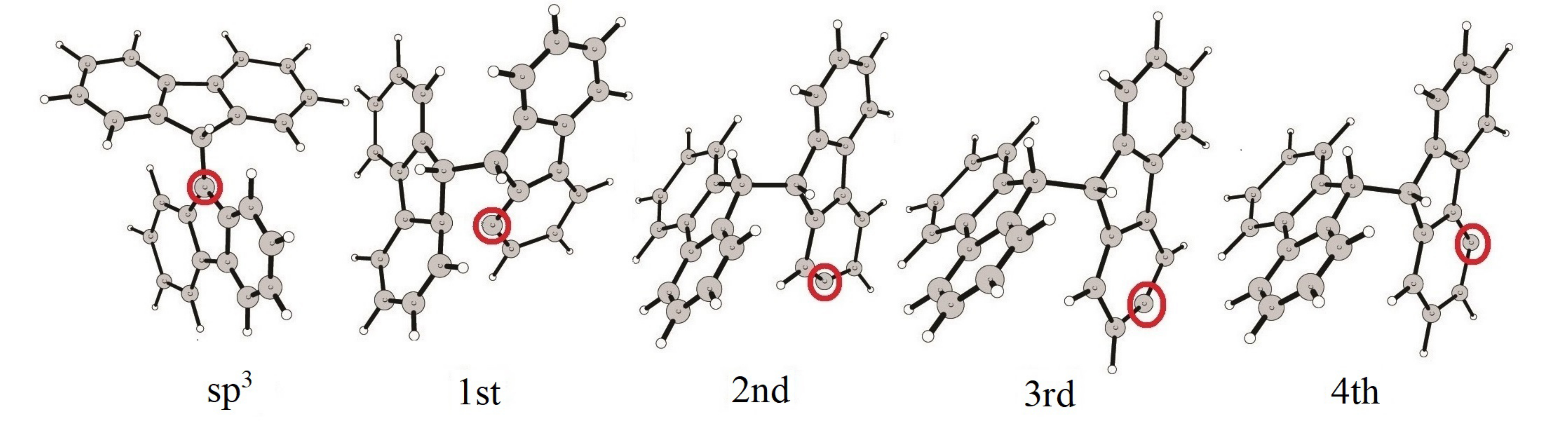}
	\caption{The possible positions in C$_{26}$H$_{18}$$^+$ from which an H atom has been lost are labeled here as sp$^3$, 1st, 2nd, 3rd and 4th, respectively.}
	\label{fig4}
\end{figure*}

\begin{figure}[t]
	\centering
	\includegraphics[width=\columnwidth]{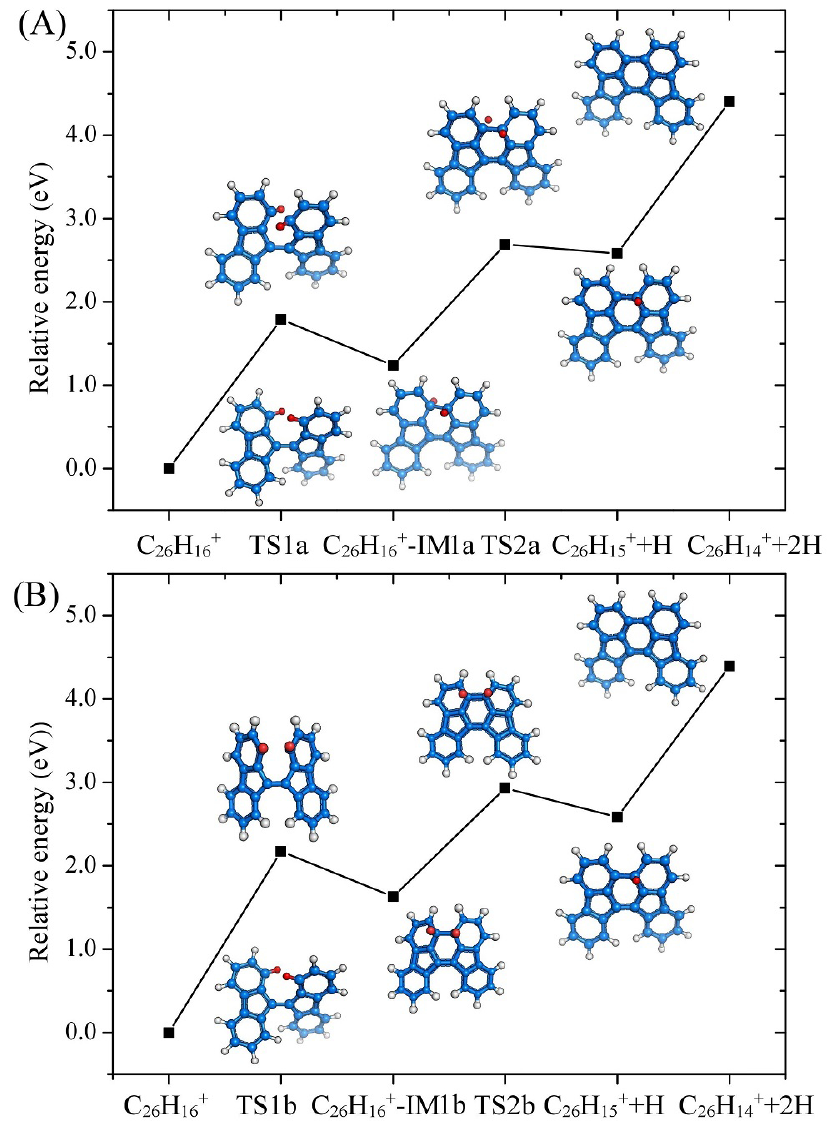}
	\caption{The reaction pathway of fluorene dimer cluster cation: panel (A) from C$_{26}$H$_{16}$$^+$ to C$_{26}$H$_{14}$$^+$ with H $+$ H loss channel; and panel (B) from C$_{26}$H$_{16}$$^+$ to C$_{26}$H$_{14}$$^+$ with the other H $+$ H loss channel. Carbon is shown in blue, nonreactive hydrogen in gray, and abstracted hydrogen in red.}
	\label{fig5}
\end{figure}

\begin{figure}[t]
	\centering
	\includegraphics[width=\columnwidth]{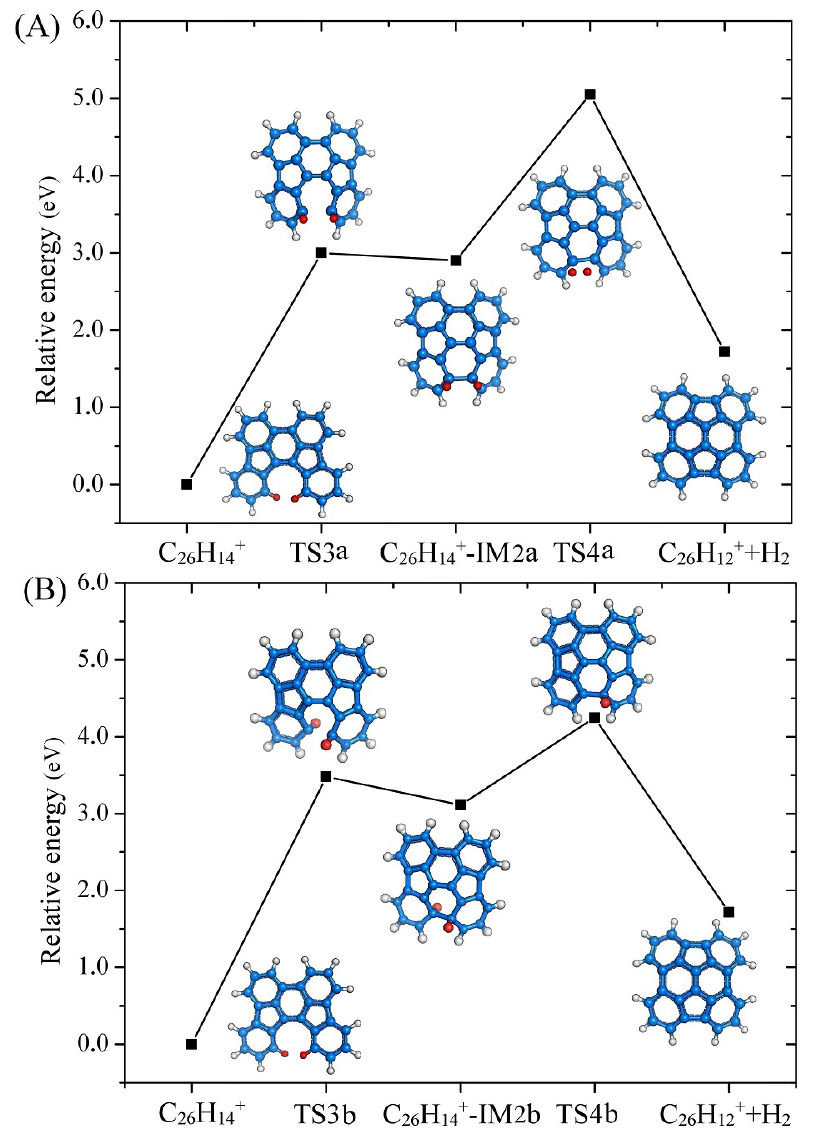}
	\caption{The reaction pathway of fluorene dimer cluster cation: panel (A) from C$_{26}$H$_{14}$$^+$ to C$_{26}$H$_{12}$$^+$ with H$_2$ loss channel; and panel (B) from C$_{26}$H$_{14}$$^+$ to C$_{26}$H$_{12}$$^+$  with H $+$ H loss channel. Carbon is shown in blue, nonreactive hydrogen in gray, and abstracted hydrogen in red.}
	\label{fig6}
\end{figure}

\section{Theoretical calculation results and discussion}
\label{sec:theoretical}

We observed a series of m/z as shown in Figures 1 and 2, and we employ quantum chemistry to link these observations to their structures and identify possible reaction pathways. As peaks in a mass spectrum can refer to more than one isomer, it is difficult to infer reaction products and pathways from the experiments alone. Therefore, the interpretation of experimental data is further investigated by density functional theory (DFT) calculations. We have used DFT to calculate the structure of the dominant fluorene dimer and trimer cations (C$_{26}$H$_{18}$$^+$ and C$_{39}$H$_{26}$$^+$). The calculations result in the optimized carbon skeletons as shown in Figure 3 (see supporting information for details).

Based on our calculations, the dimer dissociation process from C$_{26}$H$_{18}$$^+$ to C$_{26}$H$_{12}$$^+$ occurs in three steps. The first step (shown in Figure 3(A)) is the dehydrogenation from C$_{26}$H$_{18}$$^+$ to C$_{26}$H$_{16}$$^+$, losing two hydrogen atoms one by one from the sp$^3$ hybridized carbon atoms of the fluorene dimer cation with a dissociation energy smaller than 3.0 eV (2.3 and 2.6 eV, respectively). Such an H-atom step-by-step dissociation, from C$_{26}$H$_{18}$$^+$ (m/z=328) to C$_{26}$H$_{17}$$^+$ (m/z=327), and from C$_{26}$H$_{17}$$^+$ to C$_{26}$H$_{16}$$^+$ (326), indicates that C$_{26}$H$_{17}$$^+$ acts as an intermediate (IM).

From our calculations, we conclude that the C-H bond dissociation plays an important role in the polymerization process. We thus compared the dissociation energies for hydrogen from different positions (corresponding to C$_{26}$H$_{18}$$^+$ to C$_{26}$H$_{17}$$^+$), as shown in Figure 4. The DFT calculations show that the dissociation energies of hydrogen are 2.3, 3.5, 5.1, 5.0 and 5.0 eV from sp$^3$, 1st, 2nd, 3rd and 4th positions, respectively. It indicates that losing a hydrogen atom from the sp$^3$ hybridized carbon atoms is much easier than losing an aromatic hydrogen. This is a common characteristic of mixed aromatic-aliphatic species \citep{jol05, chen2015, tri17, cas18,west18}. As such, theoretical calculations confirm that the pathway in Figure 3(A) (losing H from the sp$^3$ position) is energetically preferable. 
	
In addition, for the first step, as shown in Figure 3, the two hydrogen atoms attached to sp$^3$ hybridized carbon atoms might be dissociated at the same time to form H$_2$ (see Figure 3(B)), rather than be lost one by one. We thus studied this H$_2$ releasing reaction pathway and found that the reaction barrier is 3.1 eV, which is much larger than the barrier for losing the 1st H atom (2.3 eV), as shown in Figure 3(A). We confirm that the C$_{26}$H$_{17}$$^+$ acts as an intermediate, in agreement with the experimental results (Figure 1). 

The second step is illustrated in Figure 5(A). After losing these two hydrogen atoms, the C-C bond between the two fluorene monomers can rotate more freely with a small energy barrier, resulting in isomerization. As shown here, the reaction barrier for the isomerization process is only 1.8 eV, much lower than needed for a dehydrogenation step. At the same time, a cyclization process is observed by C-C bond formation to produce the intermediate structure, IM1a. As the six-membered carbon ring is formed, the additional H on the over-coordinated carbon (sp$^3$) atom is lost. Specifically, our calculation shows that the reaction barrier, needed to overcome to lose the first hydrogen atom is only 1.45 eV (from IM1 to C$_{26}$H$_{15}$$^+$ $+$ H) and it takes 1.8 eV for the second step (from C$_{26}$H$_{15}$$^+$ to C$_{26}$H$_{14}$$^+$ $+$ H). The two-step dissociation process will generate C$_{26}$H$_{15}$$^+$, which is also reflected in the mass spectrum with high intensity. Our calculations demonstrate that both dehydrogenation processes are much faster than that in the first step from C$_{26}$H$_{18}$$^+$ to C$_{26}$H$_{16}$$^+$ due to their lower dissociation energies. During this step, the distorted fluorene dimer cation becomes a large planar PAH with 2 five- and 5 six-carbon rings. This mechanism shows many similarities to that recently studied for the photochemistry of pyrene-based clusters \citep{zhen2018}.

In the third step, two more hydrogen atoms are lost from C$_{26}$H$_{14}$$^+$ to produce C$_{26}$H$_{12}$$^+$ (Figure 6(A)). C$_{26}$H$_{14}$$^+$ isomerizes, overcoming the transition state (TS) with a barrier of 3.0 eV to form IM2a. Compared to the isomerization process in the second step, the barrier in this final step is much higher, as isomerization of a large planar PAH is more difficult than for a distorted one. Along this isomerization step, the planar PAH is forced to form a bowl structure, composed of five- and six-membered carbon rings, which can be easily recognized as a building block of a fullerene. Similar to the second step, two hydrogen atoms are lost in the cyclization process, as H$_2$ with a reaction barrier of 2.2 eV. Because of the low barrier for the H$_2$ loss channel, the reaction pathway through C$_{26}$H$_{13}$$^+$ to C$_{26}$H$_{12}$$^+$ is energetically unfavorable. This is in line with the experimental observation that the C$_{26}$H$_{13}$$^+$ (m/z=325) mass peak only is found with rather low signal strength.

To extend on this, the competing reaction pathways of the second and third dehydrogenated processes are shown in Figure 5(B) and Figure 6(B), respectively. It is clear that the dehydrogenation of two hydrogen atoms may either take place on the same side of the carbon plane or at opposite sides. For C$_{26}$H$_{16}$$^+$ to C$_{26}$H$_{14}$$^+$, as shown in Figure 5(B), the two hydrogen atoms will be pushed to the same carbon plane, and then the hydrogen atoms are dissociated step by step. However, the reaction barrier for the first step is higher than the reaction pathway in Figure 5(A) by about 0.4 eV, so the Figure 5(A) pathway is more energetically favorable. In a similar way, for the reaction of C$_{26}$H$_{14}$$^+$ to C$_{26}$H$_{12}$$^+$, as shown in Figure 6(B), the two hydrogen atoms will be pushed to the different carbon plane, and then the hydrogen atoms are dissociated step by step. As we can see, the barrier of H + H dissociating pathway (Figure 6(B)) is higher than that of H$_2$ releasing pathway in Figure 6(A) by about 0.5 eV, so the Figure 6(A) pathway is more energetically favorable.

The photo-dissociation process from C$_{26}$H$_{18}$$^+$ to C$_{26}$H$_{12}$$^+$ outlined here is the cumulative result of dehydrogenation and isomerization steps, in which the 3D fluorene dimer cation becomes planar and then transfers into a bowl structure. As discussed in supporting information, during this process, the bonds that are present become more conjugated. Figure 1 also shows the differential spectra. We have integrated the “loss” and “gain’’ peaks in the difference spectra that without consider the lower masses, and these nearly balance. Hence, H-loss is the dominant channel and intramolecular dissociation into the two monomers is at most a very minor channel. This is somewhat surprising as DFT calculations yield that the CC bond energy linking the two monomers is only 1.4 eV, which is less than the sp$^3$ H-loss channel (2.5 eV). Once a second covalent bond has been made linking the two monomers, the ”dimer” has become planar and is in essence a large PAH. It is well known that H-loss dominates C-loss for large PAHs \citep{eke98,zhen2014a,cas18,west18}. To fluorene trimer cation, negative balance became clear in Figure 2. So, there is an efficient dissociation of fluorene trimers and likely of other larger clusters, the efficiency of polymerization toward fragmentation need be considered. The dissociation of these large clusters by the laser pulses likely generates different fluorene dimers. 

For the fluorene trimer cation, the reaction pathway and energy barrier are very similar but more complex compared to the dimer case. Here, we focus on the dehydrogenation reaction pathways shown in Figure 7, with the stable structures from C$_{39}$H$_{26}$$^+$, m/z=494, to C$_{39}$H$_{20}$$^+$, m/z=488. As discussed for the dimer cation, first the H-atoms on the sp$^3$ hybridized carbon are dissociated to form C$_{39}$H$_{24}$$^+$. Subsequently, dehydrogenation and isomerization induce large PAH formation (C$_{39}$H$_{22}$$^+$, part of the molecule becomes planar, while the other part starts bowl-forming). From C$_{39}$H$_{22}$$^+$, there are two different pathways of continued dehydrogenation: one follows a bowl-forming process (1), the other retains the 3D structure of fluorene structures (2). Both end products have the same mass: C$_{39}$H$_{20}$$^+$, m/z=488. The first process (1) is very similar to the photo-induced bowl-forming process described in detail for the fluorene dimer cation. Part of the resulting structure can be considered as a fragment of a functionalized fullerene. Different from the quadrilateral-carbon ring attached on the fullerene surface obtained in \citep{dun13}, the three-carbon ring obtained here provides a possible starting point for new molecules in the ISM. In the second pathway, the resulting structure has a spiral-like geometry, which has potential for long PAH chain growth. We mention that in the dehydrogenation process from C$_{39}$H$_{26}$$^+$ to C$_{39}$H$_{20}$$^+$, six hydrogen atoms are lost; due to the multiple aromatic/aliphatic bonds, the H-loss order might be sequential or taking place at the same time. This might be the reason that the dehydrogenation behavior of the fluorene trimer cation does not have the readily identifiable odd-even pattern as seen in the dimer case (or for other previously studied PAHs). In addition, in Figure 2, C$_2$H$_2$/CH loss and dehydrogenation photo-products are observed in the range of m/z=448-484. We observe masses corresponding to C$_{38}$H$_m$$^+$ and C$_{37}$H$_n$$^+$ with $m, n=[17, 25]$. We emphasize that these ions are not fully dehydrogenated and no pure carbon clusters are produced.

\begin{figure}[t]
	\centering
	\includegraphics[width=\columnwidth]{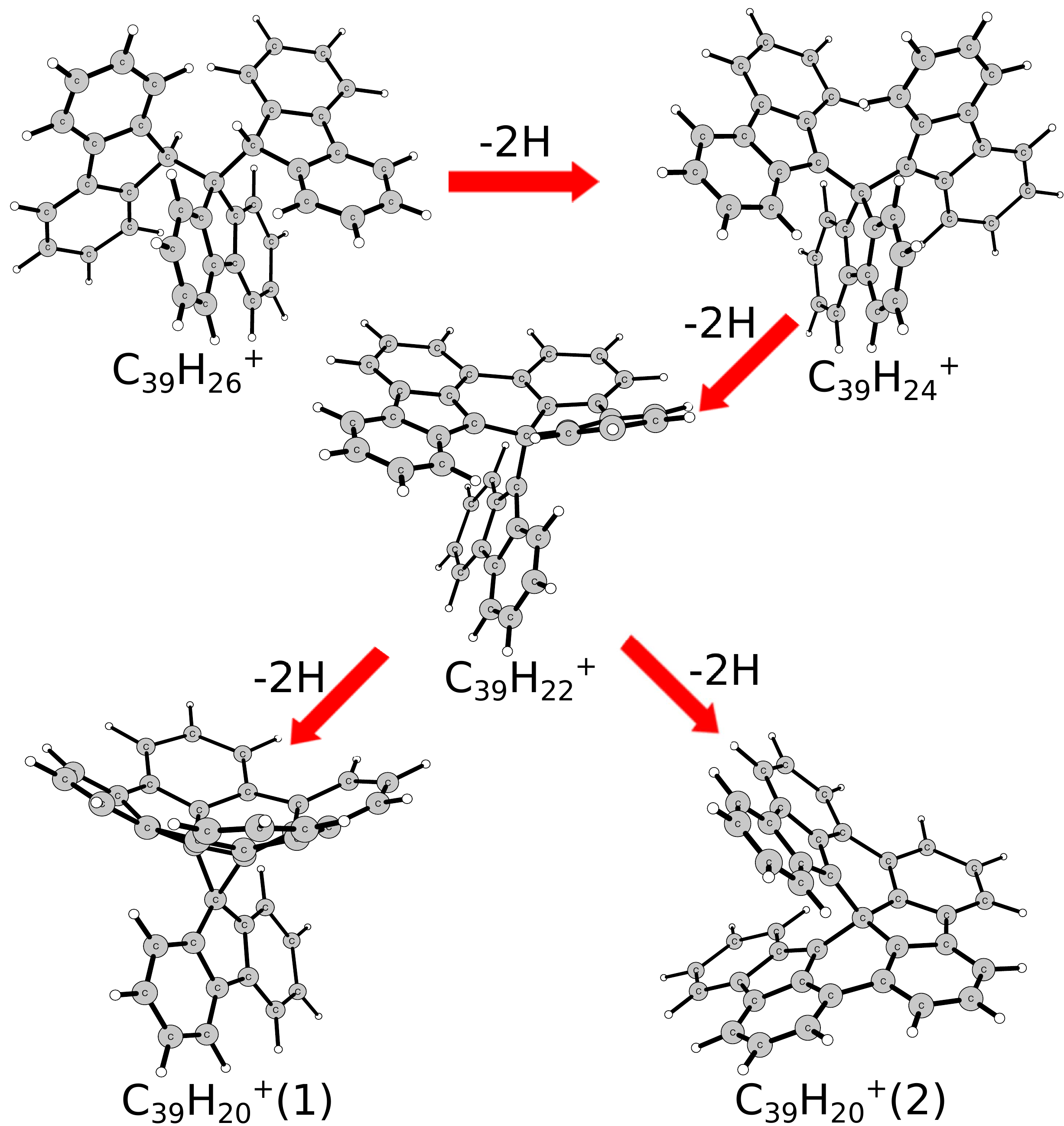}
	\caption{The dehydrogenation process of fluorene trimer cluster cation: from C$_{39}$H$_{26}$$^+$, to C$_{39}$H$_{24}$$^+$, to C$_{39}$H$_{22}$$^+$, to C$_{39}$H$_{20}$$^+$. Carbon and hydrogen atoms are shown in gray and white, respectively.}
	\label{fig7}
\end{figure}

\section{Astronomical implications}
\label{sec:discussion}

In our experiments, we employ an ion trap to isolate the covalently bonded clusters and study their subsequent evolution under irradiation. This evolution is driven by the high internal energy attained by photon absorption. Rapid internal conversion followed by intramolecular vibrational redistribution leaves the species high vibrationally excited in the ground electronic state \citep{tie08,PAHsUniverse}. Cooling occurs through a competition between vibrational emission in the IR and fragmentation (mainly H-loss for these species) \citep{mon13}. While the details of these processes (e.g., energy barriers, IR relaxation rates) will depend on the charge state, the evolution of neutrals will parallel that of ions and hence our experiments will be of general relevance. Further experiments will be needed to address the details of the evolution of neutral clusters.

As discussed in the introduction, observational and theoretical studies support the presence of PAH clusters in space \citep{rapacioli05a, rhee07}. Observations show that some 3 \% of the elemental carbon is locked up in PAH clusters deep in PDRs, such as NGC 7023 \citep{rapacioli05a, tie08}. For comparison, C$_{60}$ locks up less than 2x10$^{-4}$ of the elemental carbon in NGC 7023 and some 10$^{-3}$ in the diffuse ISM \citep{cam16, ber17}. Hence, from an abundance point of view, the formation of fullerenes through photoprocessing of PAH clusters is feasible.  In principle, as outlined in \citet{zhen2018}, photolysis "pure" PAH clusters could lead to large PAHs which then could photochemically evolve to fullerene in a top-down fashion \citep{zhen2014b}. Whether processing of small PAHs with pentagons can facilitate the formation of fullerenes as suggested by the experiments and calculations presented here depends on whether interstellar PAHs contain pentagons. 

It is not known whether interstellar PAH have pentagons in their structure. There is spectroscopic support for the presence of carbonyl groups in the PAH family \citep{tie08} and we have shown in a previous study that PAH-quinones will readily evolve photo-chemically towards pentagon-containing structures before they start losing H-atoms \citep{chen2018}. Laboratory studies have revealed that quinones are readily made by photolysis of PAHs in ices \citep{ber03}.

Experiments and theory have shown that fullerenes could form from the photo-chemical processing of large PAHs \citep{berne12, zhen2014b}. In this scenario, large PAHs originate from the ejecta of dying stars and then are then broken down by UV photolysis in the ISM into more stable species, such as C$_{60}$. In this study we have demonstrated that the photochemical evolution of charged PAH clusters, containing pentagon structures, leads in a natural way to curvature, which is a first and essential step towards fullerene formation. It also opens up bottom-up chemical scenarios where small PAHs grow through cluster formation and photolysis to fullerenes bypassing the large PAH intermediaries. This results adds to a few other studies that addressed the formation of non-planar PAHs, such as through the photochemical processing of functionalized PAHs \citep{de17, chen2018}. While full analysis has to await kinetic studies, we emphasize that the reservoir of PAH clusters is ample to supply C$_{60}$ in the ISM; The more, the caged C$_{60}$, as well as larger fullerenes, is expected to be much more stable than PAH clusters in the ISM \citep{berne2015}.

H-loss only happens after absorption of photons that raise the internal energy to $\sim$10$-$20 eV depending on the size of PAHs \citep{zhen2014a}. The same holds for the ISM. PAHs of the size that are relevant for space require multi-photon absorption before fragmentation will occur \citep{berne2015, and15}. We stress here that the derived energy barriers (from 1.5 eV up to 3.0 eV) for all the dehydrogenation and isomerization steps are smaller than the dehydrogenation energy barrier of pure PAHs ($\sim$ 4.5 eV) \citep{chen2015}. So, like in HI region (5.0 $<$ E $<$ 13.6 eV), the energy of a single photon is capable of achieving the photo-dissociation process of small PAH clusters (e.g., fluorene dimer cluster with 3.0 eV energy barrier). For clusters consisting of larger PAHs, absorption of multiple photons is required. While this will be a rare process, given the long timescales involved in evolution of the ISM, this can still be of importance \citep{mon14}. 

The present study shows results on a proto-typical laboratory example. As stated before, the presence of PAH cluster ions in space is still under debate. The detection of a specific PAH would be a major step forward and this links again to the work presented here. The non-planar species will be polar and should be detectable by radio astronomy, as the bowl-forming introduces a dipole moment. Emission by such species will contribute to the anomalous microwave emission \citep{lag03, ysa10}. In view of the large partition function of such large, non-planar molecules, unambiguous detection of individual species will be very challenging. First attempts in this direction, searching for the small, corannulene bowl, were unsuccessful \citep{lov05,pill09}. With the bowling concept introduced for bisanthenequinone cation \citep{chen2018} and extended in this study for fluorene clusters, a new class of PAH derivatives may become within range.

\section{Conclusions}
\label{sec:conclusions}

Combining experiments with quantum chemical calculations, we have presented evidence for the dehydrogenation and isomerization process of charged covalently bonded fluorene dimers and trimers cluster under laser irradiation, which shows that after H-loss, the clusters will isomerize and aromatize, leading to bowled and curved structures. Under the assumption that PAH-cluster ions play an important role in the ISM, this process may offer a way to form species that might act as building blocks of cages and fullerenes.

\acknowledgments

This work is supported by the Fundamental Research Funds for the Central Universities and from the National Science Foundation of China (NSFC, Grant No. 11743004 and Grant No. 11421303). Studies of interstellar chemistry at Leiden Observatory are supported through advanced-ERC grant 246976 from the European Research Council, through a grant by the Netherlands Organisation for Scientific Research (NWO) as part of the Dutch Astrochemistry Network, and through the Spinoza premie. HL and AT acknowledge the European Union (EU) and Horizon 2020 funding awarded under the Marie Sklodowska-Curie action to the EUROPAH consortium, grant number 722346. YS thanks the support of Science and Technology Development Program of Henan province (172102310164). TC acknowledge Swedish Research Council (Contract No. 2015-06501) and Swedish National Infrastructure for Computing (Project No. SNIC 2018/5-8).

\end{document}